\documentclass[%
 reprint,
 amsmath,amssymb,
 aps,
]{revtex4-2}

\usepackage{graphicx}
\usepackage{dcolumn}
\usepackage{bm}
\usepackage{color}

\begin{document}

\preprint{APS/123-QED}

\title{Indentation of an elastic arch on a frictional substrate: Pinning, unfolding, and snapping}

\author{Keisuke Yoshida}
\author{Hirofumi Wada}
\affiliation{Department of Physical Sciences, Ritsumeikan University, Kusatsu, Shiga 525-8577, Japan}

\date{February 23, 2024}
\begin{abstract}
In this study, we investigate the morphology and mechanics of a naturally curved elastic arch loaded at its center and frictionally supported at both ends on a flat, rigid substrate. 
Through systematic numerical simulations, we classify the observed behaviors of the arch into three configurations in terms of the arch geometry and the coefficient of static friction with the substrate.
A linear theory is developed based on a planar elastica model combined with Amontons--Coulomb’s frictional law, which quantitatively explains the numerically constructed phase diagram.
The snapping transition of a loaded arch in a sufficiently large indentation regime, which involves a discontinuous force jump, is numerically observed. The proposed model problem enables a fully analytical investigation and demonstrates a rich variety of mechanical behaviors owing to the interplay among elasticity, geometry, and friction. 
This study provides a basis for understanding more common but complex systems, such as a cylindrical shell subjected to a concentrated load and simultaneously supported by frictional contact with surrounding objects.
\end{abstract}

\maketitle

\section{Introduction}
Designing the proper geometry of a thin structure is crucial for realizing its functionality, and has become a central topic in the field of soft matter mechanics~\cite{vella2019buffering, HOLMES2019118}.
For example, by choosing suitable geometrical and structural arrangements for the assembly of thin strips and harnessing their elastic forces, various curved (pseudo) surfaces and their transformations have been achieved~\cite{liu2020tapered, baek2021smooth, Jones2023}.
Typically, the fundamental building blocks for such advanced materials are intrinsically straight and/or uniformly curved rods, ribbons, and plates. 
The intrinsic curvature of a slender object introduces novel types of elastic instabilities and morphologies, which may find various applications, such as mechanical metamaterials~\cite{bertoldi2017flexible}, jumping soft robotics~\cite{rus2015design, chen2020soft}, and bio-inspired actuators~\cite{wei2022bioinspired}. 

These geometrical structures are subjected to various boundary conditions, depending on how they are mounted onto an entire physical system.
In some cases, the relevant boundary conditions are not given a priori but rather in a state-dependent manner~\cite{BIGONI2015368, Bigoni2022}.
For instance, when a wire is inserted into a narrow channel, it is resisted by the frictional forces exerted from the inner side of the wall~\cite{Benson1982, liu2013effect}.
As the compressive force increases, the wire may buckle and eventually jam with undesired pressure applied to the wall~\cite{plaut1999deflections, roman2002postbuckling, chai2002post}, which is a potentially fatal problem in endoscopic surgery applications.
Therefore, the morphologies of thin structures in contact with the surrounding objects have been extensively studied at various levels of complexity, ranging from the buckling of a single strip on a flat surface~\cite{sano2017slip} to crumpled sheets during packing~\cite{donato2003scaling, boue2006spiral, witten2007stress, cerda1998conical, stoop2008morphological, alben2022packing}.
Notably, recent studies have also uncovered functional advantages of frictional contact, such as the force amplification in a capstan ~\cite{ghosal2012capstan, grandgeorge2022elastic}, elastic knots~\cite{PhysRevLett.99.164301}, interleaved plates~\cite{PhysRevLett.116.015502, PhysRevLett.126.218004}, and twisted yarn~\cite{PhysRevLett.128.078002}, as nonreciprocal mechanical responses in snap fits~\cite{yoshida2020mechanics} and grasping~\cite{yang2021grasping} and as energy absorption devices~\cite{fu2019programmable}.
However, the interplay among natural curvature, elasticity, and contact mechanics, and the consequences of their potential applications, have not been fully explored. 

\begin{figure}[b]
	\centering	
	\includegraphics[width=0.999\linewidth]{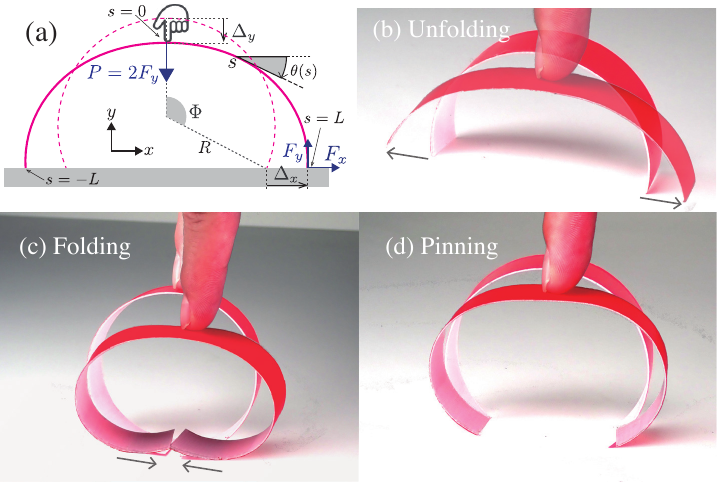}
	 \caption {Naturally curved elastic strip, which we call ``arch,'' loaded at its center on a frictional substrate.
    (a) Definition of the coordinate system and variables in our theoretical analysis. 
    (b)--(d) Photographs of the three characteristic configurations, i.e., unfolding, folding, and pinning, of a naturally curved plastic film ribbon, for illustration purposes only.
	 }\label{Fig1}
\end{figure}

To highlight the role of the intrinsic geometry and frictional contact in the mechanics of a thin structure, we propose an investigation of the simplest possible system: an elastic arch on a rigid, flat frictional surface (see Fig.~\ref{Fig1}).
Such a uniformly curved elastic strip can act as a leaf spring~\cite{frisch1962flexible} and exhibit an intriguing mechanical response that is highly dependent on its configuration. 
In addition to its geometry-dominated nature, the arch can exhibit a rich variety of force responses against indentation at its center when friction acts to resist the tangential motion (i.e., slipping) of the two ends on the surface. 

Most studies on the planar elastica problem have focused on solving the nonlinear boundary value problem, either analytically or numerically, to address its large deformation behaviors.
In this approach, transcendental equations involving elliptical integrals must be addressed~\cite{frisch1962flexible,nordgren1966finite, benson1981deformation}.
As there are generally multiple solutions to these nonlinear equations, we must also identify which solution is an energy minimizer~\cite{patricio1998elastica}.
A uniform intrinsic curvature is equivalent to a pre-moment applied externally to an intrinsically straight elastica, which makes it difficult to find such a physical solution analytically~\cite{frisch1962flexible}. 
Instead of employing a rigorous but implicit analytical approach that relies heavily on elliptic functions, we opt for a linear relationship between the force and displacement in a curved elastica by focusing on a small-amplitude deflection regime~\cite{conway1956nonlinear, Benson1982, timoshenko2009theory}. 
For example, the linear mechanical response of a curved elastic beam has been examined previously to elucidate the climbing mechanism of gecko setae~\cite{yamaguchi2009microscopic,zhao2009role}.

In this study, we consider the indentation of an elastic arch placed on a friction substrate [Figs.~\ref{Fig1} (a)]. 
We generalize the aforementioned small-deflection theory to an intrinsically curved elastic strip within the framework of the planar elastica theory. 
The predictions of our analytical theory are corroborated using systematic numerical simulations. 

The remainder of this paper is organized as follows.
In Sec.~\ref{sec_Sim}, numerical simulations are employed to investigate the deformation of an elastic arch loaded at its center on a frictional substrate~\cite{sano2017slip, yoshida2020mechanics}.
We identify three distinct behaviors, which we classify into folding, unfolding, and pinning phases, based on the static friction coefficient $\mu$ and the opening angle $2\Phi$ of the arc (see Fig.~\ref{Fig1} (a)). 
In Sec.~\ref{sec_Theory}, we develop an analytical theory based on Kirchhoff’s elastic rod theory~\cite{audoly2010elasticity} combined with a dry friction law~\cite{popov2010contact} and establish a linear response formula that quantitatively explains the mechanical phase diagram. 
In Sec.~\ref{sec_NL}, we numerically address various morphological transitions observed in a highly nonlinear deformation regime.
In Sec.~\ref{sec_Dis_Con}, we summarize our main results and discuss potential avenues for future research.

\section{Numerical simulation} \label{sec_Sim}
\subsection{Set-up of the simulation}
We consider the planar deformation of a uniform elastic arch of total length $2L$ and opening angle $2\Phi$. See Fig.~\ref{Fig1} (a).
In this study, we consider $L$ and $\Phi$ as the basic geometric parameters, and the arch is assumed to have a uniform intrinsic curvature with a radius of curvature of $R=L/\Phi$.
Gravity is ignored in this study.

The basic numerical method has been previously described~\cite{yoshida2020mechanics, sano2017slip}. 
Several key aspects of the simulation method are described briefly in this section.
A continuous elastic arch is discretized into a chain of $N=31-51$ nodes in which neighboring nodes are connected with sufficiently stiff springs of nearly constant length $a$.
The force fields are calculated from properly defined stretching and bending elastic energies~\cite{yoshida2020mechanics, sano2017slip}.
The self-avoidance of the chain is considered by the repulsive potential, modeled as a harmonic potential with a sufficiently large modulus.
The penetration of the chain into a flat rigid substrate is also prevented using the same type of potential. 
The tangential interaction of the chain nodes with the surface is modeled according to the Amontons--Coulomb law~\cite{popov2010contact}.
The position of each node evolves according to Newton’s equation of motion with a small damping parameter that ensures numerical stability.
We use the Euler iteration method to numerically integrate appropriately rescaled dynamical equations with a nondimensional time step, which is typically $0.05$.
The output values are calculated every $10^5-10^6$ steps, and the total simulation time is $10^7-10^8$ steps.

In its initial configuration, the discretized arch is semicircular and located just above a rigid flat substrate.
We impose a downward displacement at the center of the arc $\Delta_y$ and measure the reaction force $P$ with increasing $\Delta_y$ quasistatically.
As soon as both ends of the arch come into contact with the surface, they move tangentially outward or inward [see Fig.~\ref{Fig1}(b) and \ref{Fig1}(c), respectively] or remain stationary [Fig.~\ref{Fig1}(d)].
To classify the configurations of the arch quantitatively, we measure the resulting outward horizontal displacement, $\Delta_x$, and define ``unfolding'' for $\Delta_x>0$, ``pinning'' for $\Delta_x=0$, and ``folding''  for $\Delta_x<0$. 
To establish a mechanical phase diagram, we explore a wide range of parameter spaces $(\Phi, \mu)$: $\Phi= 20^\circ-178^\circ$ and $\mu=0.1-1.2$.

\begin{figure*}[]
	\centering	
	\includegraphics[width=.9\linewidth]{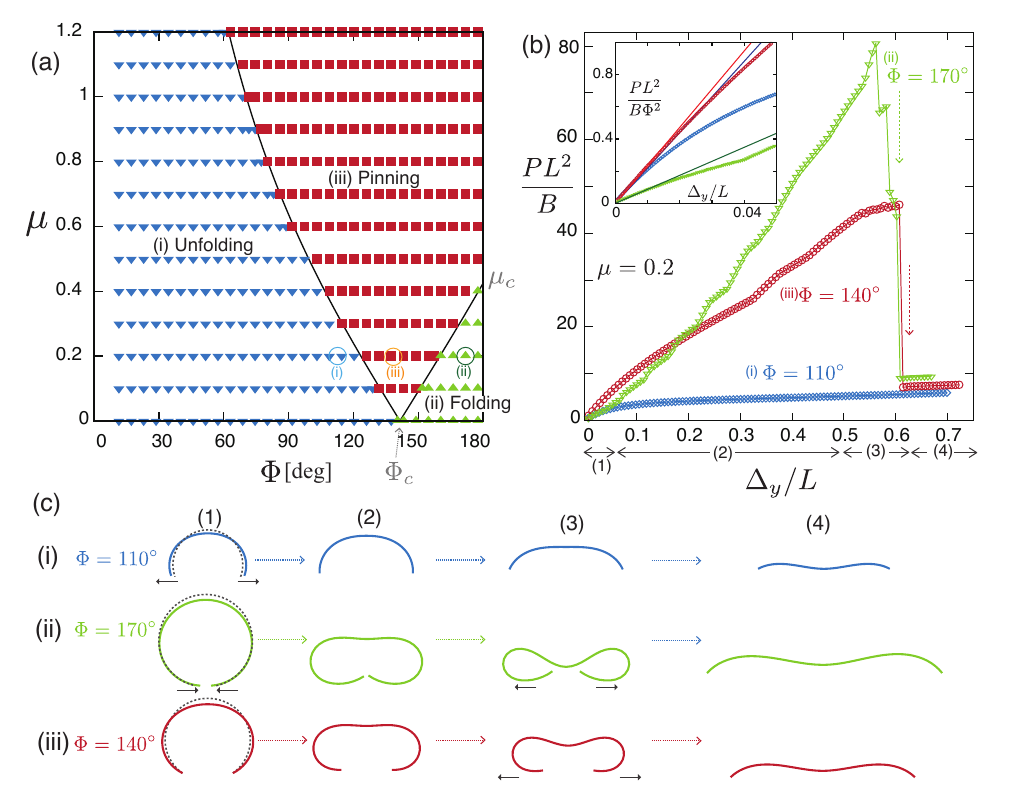}
	 \caption {Phase diagram, representative force--displacement curves, and configurations.
  (a) Numerically constructed phase diagram of the ($\Phi, \mu$) space.
  The blue, green, and red symbols represent the (i) unfolding, (ii) folding, and (iii) pinning phases, respectively.
  The solid lines represent our analytical predictions; see the main text.
  (b) Rescaled indentation force $PL^2/B$ as a function of the rescaled vertical displacement at the center of the arch, $\Delta_y/L$, for $\mu=0.2$.
  The inset displays the same data focusing on the small indentation regime, $\Delta_y/L < 0.05$.
  Note that the vertical axis in the inset shows $PL^2/B\Phi^2$ because our linear theory is developed under the basic assumption given by $PL^2/(B\Phi^2)\ll 1$ (see the main text).
  (c) Representative snapshots of an arch for increasing $\Delta_y/L$, obtained from the numerical simulations; (1)--(4) correspond to those shown in the horizontal axis in (b). 
  The corresponding locations of (i)--(iii) on the phase diagram are also shown in (a).
  }\label{Fig2}
\end{figure*}

\subsection{Results} \label{sbsc_results}
The results of a systematic numerical investigation in the low-force regime are presented in Fig.~\ref{Fig2} (a).
We observe three distinct mechanical phases depending on the parameters $(\Phi, \mu)$. 

First, we focus on the frictionless case.
See the $\mu=0$ line in Fig.~\ref{Fig2} (a).
For a shallow arch ($\Phi<\Phi_c$), where $\Phi_c$ is determined as follows, the two ends slip outward ($\Delta_x>0)$ as soon as the arch is loaded at its center.
We define this phase as ``unfolding.''
In contrast, for a deep arch, i.e., $\Phi>\Phi_c$, the two ends initially slip inward ($\Delta_x<0$); we define this phase as ``folding.''
The boundary between the folding and unfolding phases for $\mu=0$ is $\Phi_c\approx 142^{\circ}$.

Next, when frictional interactions occur between the arch and the substrate, the phase diagram bifurcates, resulting in the emergence of the third phase, which we define as ``pinning.''
See the $\mu>0$ region in Fig.~\ref{Fig2} (a).
In the pinning phase, both ends of the arch remain immobile during the indentation.
In Fig.~\ref{Fig2} (a), the pinning phase expands as $\mu$ increases and prevails in the diagram for a sufficiently large friction. 
In addition, the folding phase disappears when $\mu>\mu_c \approx 0.42$.
In the following section, we develop an analytical argument to corroborate these numerical findings and show that $\Phi_c$ and $\mu_c$ are universal numbers in the sense that they are independent of any material parameter, such as the Young's modulus of the arch.

In Fig.~\ref{Fig2}(b), the typical force vs. displacement curves for the three phases are shown for $\mu=0.2$. 
In the case of unfolding [Fig.~\ref{Fig2}(b) (i)], the force increases linearly with the vertical displacement $\Delta_y (\ll L)$.
With increasing vertical displacement, the force increases but deviates from the linear slope, whereas the two ends continue to slip outward [Fig.~\ref{Fig2}(b–i) and Fig.~\ref{Fig2}(c-i)].
In the case of folding [Fig.~\ref{Fig2}(b) (ii)], the force initially exhibits a linear increase with the vertical displacement, but then increases rapidly because of self-avoiding interactions [Fig.~\ref{Fig2}(b-ii)].
Note that the force curve shows irregular variations owing to the frictional contact between the arch and the susbstrate [Fig.~\ref{Fig2}(c-ii)].

Note also that the diagram in Fig.~\ref{Fig2} (a) is constructed based on the response of the arch in the low-force regime.
In the pinning phase in Fig.~\ref{Fig2} (a), the two ends remain pinned (i.e., $\Delta_x=0$) in the low-force regime.
However, when the arch is strongly squeezed [Fig.~\ref{Fig2}(b-iii) and \ref{Fig2}(c-iii)], 
the two ends suddenly snap outward, and the loading force $P$ decreases discontinuously to reach the same magnitude as that in the unfolding phase [See Fig.~\ref{Fig2}(b-i)].

\section{Theoretical analysis} \label{sec_Theory}
\subsection{Set-up of the analysis}
Focusing on the linear regime characterized by a sufficiently small vertical displacement $\Delta_y\ll L$, we rationalize the above numerical findings in the framework of the inextensible planar elastica theory~\cite{frisch1962flexible, audoly2010elasticity}.
In the two-dimensional problem, the position of the centerline of the arch is represented as ${\bm r}(s)=(x(s), y(s))$, where $s$ is defined as the arc length parameter $s\in [-L, L]$.
We assume that the center of the arch is at $s=0$, with the left and right ends at $s=-L$ and $L$, respectively.
In addition to the Cartesian coordinate representation, we define the variable $\theta(s)$ to represent the angle between the tangent line at $s$ and the horizontal axis, which is suitable for considering the inextensibility conditions implied in the elastica model.
From Fig.~\ref{Fig1} (a), $dx(s)/ds=\cos\theta(s)$ and $dy(s)/ds=-\sin\theta(s)$.
Assuming a symmetric deflection of the arch about $s=0$ as observed in our numerical simulations, we consider the deformation of only half of the arch, that is, segment $0 \leq s \leq L$.
This indicates that the arch at $s=0$ is clamped horizontally, that is, $\theta(0)=0$.

A simple geometric consideration indicates the following relationship. 
\begin{align}
\int_0^{L} \cos\theta(s) ds &=\frac{\sin\Phi}{\Phi}L+\Delta_x, \label{LateralConst} \\
\int_0^{L} \sin\theta(s) ds&=\frac{1-\cos\Phi}{\Phi} L -\Delta_y. \label{VerticalConst}
\end{align}

When a concentrated downward force $P$ is applied at $s=0$, the substrate exerts a reaction force $(F_x,F_y)$ on the (right) edge $(s=L)$.
The overall vertical force balance is $P=2F_y$.
The elastica equilibrium equation~\cite{frisch1962flexible, audoly2010elasticity} is given by
\begin{align}
B\theta''(s)-F_x\sin\theta(s) - F_y\cos\theta(s)=0, 	\label{eq:Elastica}
\end{align}
where the prime symbol $()^{\prime}$ represents the derivative with respect to $s$, and $B$ is the bending modulus.
In addition, we assume moment-free boundary conditions at both ends~\cite{sano2017slip, yoshida2020mechanics}, implying $\theta'(L)=\Phi/L$.
We also assume that there is no inflection point in the shape of our elastica considered here.
Integrating Eq.~(\ref{eq:Elastica}) and using the boundary conditions at $s=L$ given above, we find that 
\begin{align}
\frac{ds}{L} =\frac{d\theta}{\Phi} 
	\left[
	1+\frac{2}{\Phi^2}\left(
    		\begin{array}{c}
      			f_x\\
     			f_y
      		\end{array}
  	\right)\cdot 
	\left(
    		\begin{array}{c}
      			-\cos\theta+\cos\theta(L)\\
     			\sin\theta-\sin\theta(L)
      		\end{array}
  	\right)
	\right]^{-1/2},
	\label{eq:1stInt:Exact}
\end{align}
where we define the nondimensional force as
\begin{eqnarray}
 f_{\alpha} &=& \frac{F_{\alpha}L^2}{B},
 \label{eq:f_alpha}
\end{eqnarray}
for $\alpha=x, y$.
Equation~(\ref{eq:1stInt:Exact}) can be integrated exactly using elliptical integrals; however, this is notoriously cumbersome for naturally curved beams~\cite{frisch1962flexible}.
Instead, focusing on a low-force regime characterized by $f_{\alpha}/\Phi^2 \ll 1$ for $\alpha=x$ and $y$, we linearize Eq.~(\ref{eq:1stInt:Exact}), and obtain
\begin{align}
\frac{ds}{L} &\approx \frac{d\theta}{\Phi} 
	\left[
	1 - \frac{1}{\Phi^2}\left(
    		\begin{array}{c}
      			f_x\\
     			f_y
      		\end{array}
  	\right)\cdot 
	\left(
    		\begin{array}{c}
      		-\cos\theta+\cos\theta(L)\\
     			\sin\theta-\sin\theta(L)
      		\end{array}
  	\right)
	\right]. \label{eq:1stInt}
\end{align}
By applying the integrals in Eqs.~(\ref{LateralConst}) and (\ref{VerticalConst}), using Eq.~(\ref{eq:1stInt}) and the condition of inextensibility, we obtain a linear relationship given by 
\begin{align}
\frac{\Delta_\alpha}{L}=C_{\alpha \beta}\frac{F_\beta L^2}{B}, 	\label{eq_LR_gene}
\end{align}
where the Greek indices run in the $x$ and $y$ directions, indicating Einstein's sum rule.
The symmetric compliance matrix $C_{\alpha\beta}$ in Eq.~(\ref{eq_LR_gene}) is defined as follows:
\begin{align}
C_{xx}(\Phi) &= \frac{1}{2\Phi^2} - \frac{\cos\Phi}{\Phi^3} \left( \frac{3}{2} \sin\Phi - \Phi \cos\Phi \right), \label{eq_C_xx} \\
C_{xy}(\Phi) &= \frac{1}{2\Phi^3} + \frac{\cos\Phi}{\Phi^3} \left( 1-\frac{3}{2} \cos\Phi - \Phi \sin\Phi \right), \label{eq_C_xy} \\
C_{yx}(\Phi) &= C_{xy}(\Phi),  \label{eq_C_yx} \\
C_{yy}(\Phi) &= \frac{1}{2\Phi^2} + \frac{\sin\Phi}{\Phi^3} \left( -2+\frac{3}{2} \cos\Phi + \Phi \sin\Phi \right).
\label{eq_C_yy}
\end{align}
The inverse of the matrix ${\bf C}$, which we write as ${\bf C}^{-1}$, is the (nondimensional) rigidity matrix that can be determined from $\Phi$ only.
The off-diagonal elements of ${\bf C}$ provide normal tangential coupling, which quantifies the lateral displacement induced by the vertical loading force.

\subsection{Effective stiffness of the arch}
\begin{figure}[]
	\centering	
	\includegraphics[width=1.\linewidth]{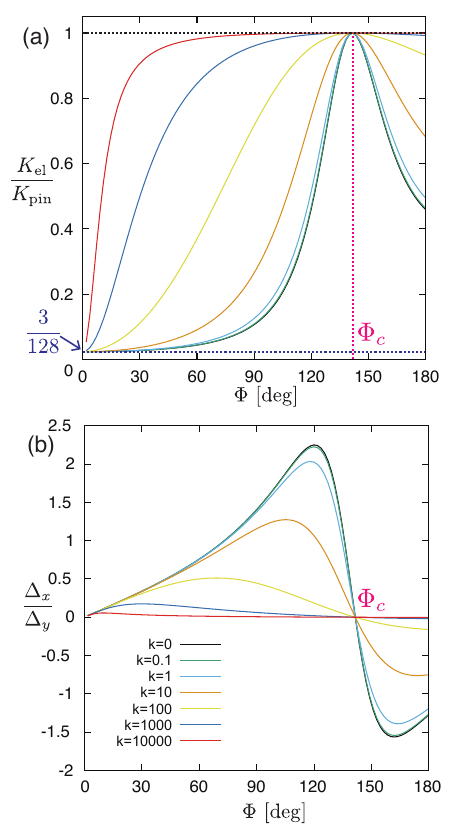}
	 \caption {Analytical results of our linear theory for various values of the boundary stiffness $k$.
    (a) Normalized effective stiffness $K_{\rm el}/K_{\rm pin}$ plotted as a function of the opening angle $\Phi$ for various $k$ [indicated in the legend of (b)], based on Eq.~(\ref{eq_K_el}). 
    (b) Ratio of the vertical ($\Delta_y$) to horizontal ($\Delta_x$) displacements plotted as a function of $\Phi$ for various $k$ (indicated in the legend), based on Eq.~(\ref{eq_DelxDely}).
	 }\label{Fig3}
\end{figure}

To characterize the mechanical properties of an elastic arch on a flat substrate, we derive Hooke's law of indentation as follows:
\begin{align}
\frac{PL^2}{B}=K(\Phi)\frac{\Delta_y}{L},	\label{LR} 
\end{align}
where $K(\Phi)$ denotes the dimensionless effective stiffness.
A specific form of $K(\Phi)$ can be obtained analytically by using Eq.~(\ref{eq_LR_gene}), once the lateral force $F_x$ is determined by the given boundary conditions.
To encompass the different conditions imposed along the $x$ direction, we first assume a simple but general boundary condition in which the end positions of the arch are constrained by linear springs with the dimensionless stiffness $k$; that is,
\begin{align}
    \frac{F_xL^2}{B} = -k\frac{\Delta_x}{L}.
    \label{F_x_sp}
\end{align}
For $k=0$, the elastic strip is free to move horizontally (on a frictionless substrate), whereas for $k\rightarrow \infty$, the ends are firmly fixed at their initial positions.
Substituting Eq.(\ref{F_x_sp}) into Eq.~(\ref{eq_LR_gene}), we obtain 
\begin{eqnarray}
    K_{\rm el}(\Phi,k) &=& K_{\rm pin}(\Phi)\frac{k+(C_{xx})^{-1}}{k+(C^{-1})_{xx}},
    \label{eq_K_el}
\end{eqnarray}
where
\begin{eqnarray}
    K_{\rm pin}(\Phi) &=& 2(C^{-1})_{yy}.
    \label{eq_Kpin}
\end{eqnarray}
We define $(C^{-1})_{\alpha \beta}$ as the $(\alpha,\beta)$-component of the inverse matrix ${\bf C}^{-1}$ and $(C_{\alpha \beta})^{-1}=1/C_{\alpha \beta}$.
$K_{\rm pin}(\Phi) = \lim_{k\rightarrow \infty} K_{\rm el}(\Phi,k)$ denotes the stiffness of the pinning boundary conditions.
In Fig.~\ref{Fig3} (a), we plot $K_{\rm el}/K_{\rm pin}$ as a function of $\Phi$ for various values of $k$ (as indicated in the legend of Fig.~\ref{Fig3} (b)).
A shallow strip ($\Phi\ll 1$) is much softer for a finite $k$ compared with the pinning case ($k=\infty$).
In particular, we find that 
\begin{eqnarray}
 \lim_{\Phi\rightarrow 0} \frac{K_{\rm el}(\Phi, 0)}{K_{\rm el}(\Phi, \infty)} &=& \frac{3}{128}.
 \label{eq:ratio}
\end{eqnarray}
The relationship between the lateral and vertical displacements can also be obtained as
\begin{align}
    \frac{\Delta_x}{\Delta_y} = -\frac{(C^{-1})_{xy}}{k + (C^{-1})_{xx}},
    \label{eq_DelxDely}
\end{align}
and are plotted in Fig.~\ref{Fig3} (b).
An indentation, that is, an increase in $\Delta_y$, induces a lateral displacement $\Delta_x$, and the magnitude of this coupling depends on the intrinsic curvature $\Phi$. 
As observed in Figs.~\ref{Fig3} (a) and (b), we have  $K_{\rm el}/K_{\rm pin}=1$ at $\Phi=\Phi_c$ irrespective of the value of $k$, at which $|\Delta_x/\Delta_y|=0$. 
Thus, $\Phi_c$ is determined by solving $(C^{-1})_{xy}(\Phi)=0$ numerically, from which we find $\Phi_c \approx 142^\circ$.
For $\Phi=\Phi_c$, the ends of the arch are virtually pinned irrespective of the value of $k$, which is purely geometric.
Note also that Eq.~(\ref{eq_DelxDely}) predicts the unfolding ($\Delta_x>0$) and folding ($\Delta_x<0$) phases for shallow $(\Phi < \Phi_c)$ and deep $(\Phi > \Phi_c)$ arches (see Fig.~\ref{Fig3} (b)), which agrees with the diagram shown in Fig.~\ref{Fig2} (a). 
Considering that the present analytical argument applies to any frictionless ($\mu=0$) substrate, we can fully explain the threshold angle $\Phi_c$ observed in the diagram given in Fig.~\ref{Fig2}.

To rationalize the numerically constructed diagram shown in Fig.~\ref{Fig2}, we consider frictional interactions with the substrate, assuming the Amontons--Coulomb friction law.
Based on the classifications shown in Fig.~\ref{Fig2}, the stiffness $K(\Phi)$ differs among the three phases. 
The stiffness of the pinning phase is expressed in Eq.~(\ref{eq_Kpin}).
In the unfolding and folding phases, the quasistatic loading assumption indicates the critical condition $|F_x/F_y|=\mu$.
Combining this with Eq.~(\ref{eq_LR_gene}), we obtain
\begin{align}
K_{\pm}(\Phi,\mu)=K_{\rm pin} \frac{(C_{xx})^{-1}}{(C^{-1})_{xx}\pm \mu (C^{-1})_{xy}},
\label{eq_Kpm}
\end{align}
where the positive and negative indices in $K_\pm$ represent unfolded and folded phases, respectively.
Note that Eq.~(\ref{eq_Kpm}) with $\mu=0$ agrees well with Eq.~(\ref{eq_K_el}), where $k=0$.
In the inset of Fig.~\ref{Fig2} (b), we plot the force law based on Eqs.~(\ref{eq_Kpin}) and (\ref{eq_Kpm}) and observe consistency with our simulation results.

\subsection{Phase boundaries}
Using the aforementioned theoretical framework, we analytically determine the phase boundaries as shown in Fig.~\ref{Fig2} (a).
First, we consider an elastic arch in a pinning configuration.
By substituting $\Delta_x=0$ in Eq.~ (\ref{eq_LR_gene}), we immediately obtain ${F_x}/{F_y}=- {C_{xy}}/{C_{xx}}$.
The critical condition $\mu=|F_x/F_y|$ is satisfied for the two-phase boundary lines shown in Fig.~\ref{Fig2} (a).
We can find that $C_{xy}>0$ for $\Phi<\Phi_c$ and the pinning--unfolding boundary line is determined by 
\begin{align}
\mu = \frac{C_{xy}}{C_{xx}}. \label{eq_PB_UP}
\end{align}
Similarly, because $C_{xy}<0$ for $\Phi>\Phi_c$, the pinning--folding boundary line can be determined as follows. 
\begin{align}
\mu = - \frac{C_{xy}}{C_{xx}}. \label{eq_PB_FP}
\end{align}
We plot Eqs.~(\ref{eq_PB_UP}) and (\ref{eq_PB_FP}) in Fig.~\ref{Fig2} (a), which is consistent with the simulation results.
The critical value of the static friction coefficient $\mu_c$, above which the folding phase disappears for all $\Phi$ values, is directly obtained from Eq.~(\ref{eq_PB_FP}) as 
\begin{equation}
 \mu_c=-\frac{C_{xy}(180^{\circ})}{C_{xx}(180^{\circ})} = \frac{4}{3\pi} \approx 0.424,
\end{equation}
This is in quantitative agreement with the numerical results shown in Fig.~\ref{Fig2} (a).
We then establish two critical parameters, $\Phi_c$ and $\mu_c$, which are independent of elastic properties, such as Young's modulus; hence, they are purely geometric.
Note that Eq.~(\ref{eq_PB_UP}) asymptotically approaches $25/(16\Phi)$ for $\Phi \ll 1$, indicating that the unfolding region decreases as $\mu$ increases, but remains for an arbitrarily large value of $\mu$.

\begin{figure}[]
	\centering	
	\includegraphics[width=1.\linewidth]{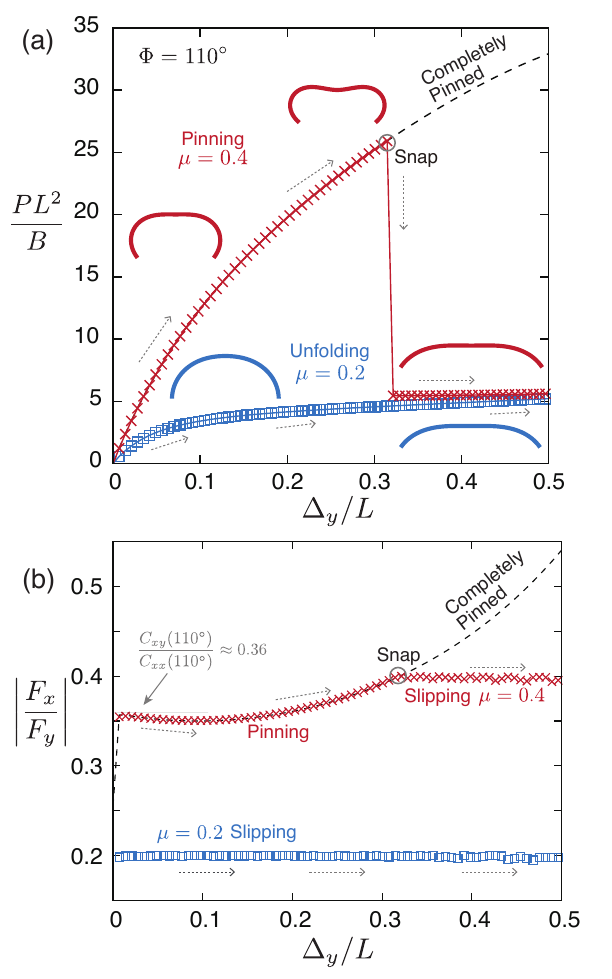}
	 \caption {Snapping behavior and underlying friction mechanism along with the arch configurations.
    (a) Rescaled indentation force $PL^2/B$ plotted as a function of the rescaled vertical displacement $\Delta_y/L$ for $\Phi=110^{\circ}$ and for $\mu=0.2$ (red, unfolding) and $\mu=0.4$ (blue, pinning).
    The insets show several representative configurations from the numerical simulations. 
    (b) Plot of $|F_x/F_y|$ as a function of $\Delta_y/L$ for the two cases shown in (a).
    The snap point indicated in the data for $\mu=0.4$ corresponds to the snap point of the significant force drop in (a).
    The dotted line is the prediction for the hinge--hinge (i.e., ``completely pinned'') boundary condition.
    }\label{Fig4}
\end{figure}

\section{Large indentations} \label{sec_NL}
When the loading displacement is sufficiently large, the mechanical response significantly changes.
Figure~\ref{Fig4}(a) shows a typical force-–displacement curve obtained from the numerical simulation.
As discussed in Sec.~\ref{sbsc_results}, for $(\Phi, \mu)=(110^\circ, 0.2)$, the curved strip unfolds upon indentation at its center, and as the displacement increases, the loading force increases more slowly than its initial linear response slope.
To understand the characteristic force behavior in terms of the strip configuration, we plot $|F_x/F_y|$ as a function of $\Delta_y/L$ in Fig.~\ref{Fig4}(b).
For $\mu=0.2$ (red symbols), the strip undergoes quasistatic slipping because $|F_x/F_y|\simeq \mu$ holds in Fig.~\ref{Fig4}(b).
By contrast, for $(\Phi,\mu)=(110^\circ, 0.4)$ (blue symbols), the strip ends are initially pinned as $|F_x/F_y| < \mu$ for a small $\Delta_y/L$. 
In fact, Eq.~(\ref{eq_PB_UP}) predicts $|F_x/F_y|=C_{xy}(110^\circ)/C_{xx}(110^\circ)\approx 0.36 < \mu\,  (= 0.4)$, which is consistent with the numerical results shown in Fig.~\ref{Fig4}(b).
Although our linear theory predicts that $|F_x/F_y|$ is independent of $\Delta_y$, beyond this linear regime, $|F_x/F_y|$ is observed to increase with increasing $\Delta_y/L$.
As soon as $|F_x/F_y|=\mu$ (= 0.4 in the present case), the strip ends snap outward and the force decreases discontinuously, as shown in Figs.~\ref{Fig4}(a) and~\ref{Fig2}(b-iii).
Beyond this transition point, the critical condition $|F_x/F_y|\approx \mu$ holds for increasing $\Delta_y/L$, whereas the magnitude of the force is of the same order as that for $\mu=0.2$ (Fig.~\ref{Fig4}(a)).
For comparison, we numerically investigate an elastic arch firmly pinned at its two ends and loaded at its center.
As observed in Fig.~\ref{Fig4} (a), the force--displacement curve is in excellent agreement with that for $\mu=0.4$ provided that $|F_x/F_y|<\mu$.
Therefore, the critical displacement for the pinning-to-slipping transition (accompanied by snapping) can be precisely predicted using this comparison.

\begin{figure}[]
	\centering	
	\includegraphics[width=1.05\linewidth]{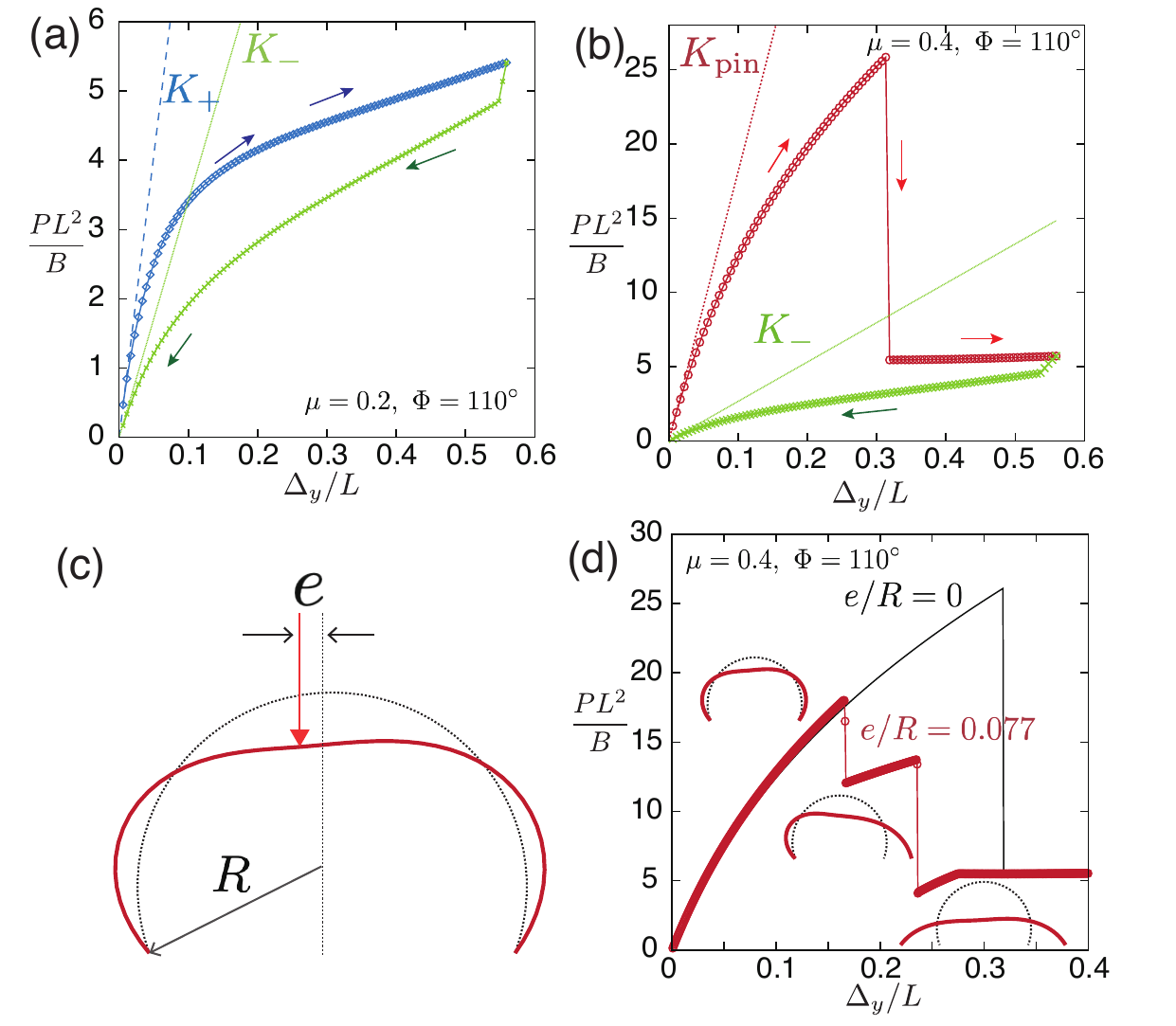}
	 \caption{Numerical simulation results on the hysteretic force response during a cycle process consisting of the forward (pushing) and backward (relaxing) indentations.
    Rescaled indentation force $PL^2/B$ plotted as a function of the rescaled vertical displacement $\Delta_y/L$ for $\Phi=110^{\circ}$, and (a) for $\mu=0.2$ (unfolding phase), and (b) $\mu=0.4$ (pinning phase).
    The dashed and solid lines represent the predictions from our linear theory. See Eq.~(\ref{eq_Kpm}). 
    (c) Schematics of off-center indentation geometry. A point of indentation is at a distance $e$ away from the center of the arch.
    (d) $PL^2/B$ plotted as a function of $\Delta_y/L$ observed in the off-center indentation of $e/R=0.077$, $\Phi=110^{\circ}$, and $\mu=0.4$, together with the representative configurations of the arch.
    }
    \label{fig:figure5}
\end{figure}

In the presence of friction, a distinctive hysteretic behavior can appear during the cyclic indentation process.
Figure~\ref{fig:figure5} shows the typical hysteresis loop observed in the unfolding phase [Fig.~\ref{fig:figure5} (a)] and pinning phase [Fig.~\ref{fig:figure5} (b)].
In the unfolding phase, as the critical condition $|F_x/F_y|\approx \mu$ always holds, the frictional force acting on the ends of the strip changes direction during the forward (pushing) and backward (relaxing) processes.
This is most evident in the difference in the linear slopes in the small-force regimes.
According to the linear theory expressed in Eq. (\ref{eq_Kpm}), the linear slope in the backward process should become $K_{+}(-\mu)=K_{-}(\mu)$, as indicated by the solid line in Fig.~\ref{fig:figure5} (a) and is in excellent agreement with the numerical data. 
In contrast, in the pinning phase, the force response is perfectly reversible {\it before} the snapping transition, which is evident, considering that the cycle process is undertaken virtually under fixed boundary conditions. 
However, a substantial hysteresis appears when the direction of the indentation changes {\it after} the snapping transition.
As the arch is considerably unfolded during snapping, the strip behavior in the backward process becomes similar to that in the unfolding phase.
In this case, both the asymmetry originating from the nature of the dry friction and the asymmetry arising from the strip configurations 
contribute significantly to the remarkable hysteretic force behavior shown in Fig.~\ref{fig:figure5} (b).

\section{Discussion and conclusion} \label{sec_Dis_Con}
By combining numerical simulations and analytical theory, we quantify the mechanical responses of an elastic arch loaded at its center and frictionally supported at both ends on a flat substrate. 
We formulate a compliance matrix for the arch against indentation, which is a function of the intrinsic geometry of the arch. 
When combined with specific boundary conditions, this allows us to obtain the effective Hooke's law.
Our theoretical framework quantitatively predicts a numerically constructed phase diagram.

We also numerically investigate the deformation behavior of the arch at large indentations. 
The arch exhibits a snapping shape transition that cannot be captured by linear response analysis.
We also show that the force drastically decreases during the snapping transition, which is potentially useful for application to efficient energy-absorbing materials~\cite{fu2019programmable, garland2020coulombic, sano2023randomly}.

This study has several possible extensions.
First, the configuration of a naturally curved elastica is mathematically equivalent to the so-called ``sticky elastica''  problem~\cite{wagner2013sticky}. 
Sticky elastica refers to the delaminated configuration of an elastic strip that initially adheres to a wet substrate or liquid surface. 
In this setup, the ends of the strip must satisfy the boundary condition $\theta'\propto 1/\ell_{\rm ec}$, where $\ell_{\rm ec}$ is often termed as ``elasto-capillary length''~\cite{roman2010elasto}.
Therefore, the proposed theoretical framework can be applied to quantitatively estimate the adhesion energy of an elastic sheet or strip on a sticky surface~\cite{roman2010elasto, elder2020adhesion, napoli2022transition}.

Second, the effects of self-weight need to be investigated to determine the configuration of a heavy structure.
This is particularly relevant to either a soft object or a structure with a sufficient weight. 
A balance between the gravitational pull and the elastic restoring force naturally sets the characteristic length scale, called the gravitobending length, which is given by
$L_{\rm gb}= (B/\rho g S)^{1/3}$, where $\rho$ is the mass density, and $S$ is the cross-sectional area of the slender rod and beam \cite{wang1986critical}.
Therefore, the relative significance of the weight to elasticity for a naturally curved beam is quantified by the nondimensional parameter $R/L_{\rm gb}$ \cite{lazarus2013contorting}, where $R$ is the intrinsic radius of curvature of the arch.
For $R/L_{\rm gb} \ll 1$, the arch deforms only slightly; however, for $R/L_{\rm gb} \gg 1$, the arch exhibits an M-shape similar to that observed in Fig.~\ref{Fig2}.
A detailed study of a gravity-driven shape transition of a naturally curved elastic line will be reported elsewhere.

Third, extending our study to thin-shell mechanics is an interesting but challenging endeavor.
The indentation behavior of a semi-cylindrical shell has been extensively studied, usually under clamped boundary conditions for lengthwise edges, where the remaining two edges are free.
The localized deformation of a cylindrical shell is inevitably accompanied by in-plane stretching~\cite{audoly2010elasticity, HOLMES2019118}, which leads to stress focusing and the creation of a pair of defects~\cite{witten2007stress, vaziri2008localized}.
These individual structures can migrate to the free edges, where they disappear, thereby reducing the stretching energy and drastically decreasing the indentation force.
At this point, the shell recovers its isometrically deformed shape, which is similar to the M-shape observed in the planar elastica model~\cite{boudaoud2000dynamics,vaziri2008localized}.
It needs to be investigated how such generic behavior is modified if the edges of the shell are supported by frictional interactions with a flat substrate.

Finally, the theoretical findings must be verified through physical experiments.
In the present study, we have investigated the symmetric deformation of an ideal arch with uniform natural curvature and elastic modulus, assuming an indentation at the center of the arch. 
However, in our preliminary experiments, the arch was often observed to undergo asymmetric buckling deformation upon indentation. 
One of the major causes of the observed symmetry breaking is the intrinsic variation in the geometric parameters, such as the thickness and natural curvature, of the fabricated physical models.
This suggests the need to increase the fabrication accuracy for the quantitative experimental verification of our analytical and numerical findings. 
Another major reason may be the accuracy of the indentation position \cite{chen2011snapping}.
To examine the significance of this, we investigate the slightly off-centered indentation of a perfectly uniform, ideally shaped, elastic arch in our numerical simulation.
In Fig.~\ref{fig:figure5} (c) and (d), we show the typical arch behavior observed for off-centeredness $e/R=0.077$, $\mu=0.2$, and $\Phi=110^{\circ}$.
The initial response in terms of both the configuration and force is almost identical to that for $e/R=0$.
However, as the indentation depth increases, the end of the indentation point snaps first, accompanied by a discontinuous drop in the force response. 
At this point, a highly asymmetric configuration is observed (Fig.~\ref{fig:figure5} (d), inset), which is not observed for $e/R=0$.
Further indentation then induces a second snapping of the remaining end, again involving a discontinuous force decrease to a value comparable to that for $e/R=0$.
Therefore, for an off-centered indentation, the unique snapping point for $e/R=0$ bifurcates into the second-step process, and the resulting force curve becomes more complicated. 
Overall, these numerical findings are consistent with those observed in our preliminary experiment.
However, a large parameter space remains to be explored for a thorough characterization of the buckling behavior of indented curved elastic strips with various levels of imperfections.
Further theoretical and experimental investigations are required to understand the complex mechanical responses of curved elastic structures supported by friction substrates.

\begin{acknowledgments}
We acknowledge financial support from JSPS KAKENHI (Grant Nos. 22H05067 and 22H01192 to HW), a Grant-in-Aid for JSPS Research Fellows (DC1, No. 21J22837 to KY), and a Sasakawa Scientific Research Grant from the Japan Science Society (No.2020-2005 to KY).
\end{acknowledgments}

\end{document}